\documentclass[aps,prl,twocolumn,superscriptaddress]{revtex4}
\usepackage{graphicx}
\usepackage{psfrag}
\usepackage{amsmath}
\usepackage{amssymb}
\usepackage{amsthm}

\begin{document}

\title{The cascades route to chaos}

\author{Evelyn Sander,
George Mason University,
Fairfax, VA, 22030, U.S.A. 
E-mail: \texttt{esander@gmu.edu}
and
James A.~Yorke,
University of Maryland,
College Park, MD 20742, U.S.A. 
E-mail: \texttt{yorke@umd.edu}}
\date{October 18, 2009 }

\begin{abstract}
  The presence of a period-doubling cascade in dynamical systems that
  depend on a parameter is one of the basic routes to chaos. It is
  rarely mentioned that there are virtually always infinitely many
  cascades whenever there is one.  We report that for one- and
  two-dimensional phase space, in the transition from no chaos to
  chaos -- as a parameter is varied -- there must be infinitely many
  cascades under some mild hypotheses.  Our meaning of chaos
  includes the case of chaotic sets which are not attractors.
  Numerical studies indicate that this result applies to the forced-damped pendulum and
  the forced Duffing equations, viewing the solutions once each period
  of the forcing.  We further show that in many cases cascades appear
  in pairs connected (in joint parameter-state space) by an unstable
  periodic orbit. Paired cascades can be destroyed or created by perturbations,
  whereas unpaired cascades are conserved under even significant
  perturbations.
\end{abstract}

\maketitle

\begin{figure}
\includegraphics[width=.5\textwidth]{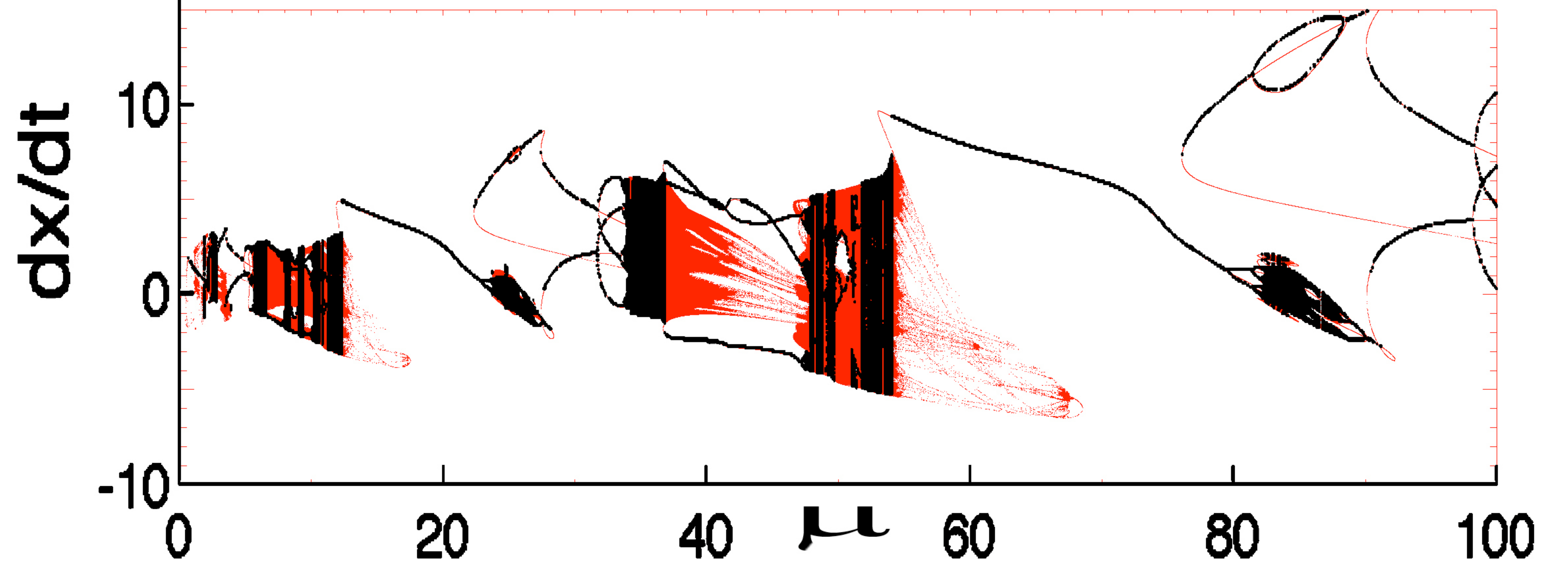}
\caption{{\bf Cascades in the double-well Duffing equation.} 
The attracting sets (in black) and periodic orbits up to period ten
(in red)
for the time-$2\pi$ map of the double-well Duffing equation: 
$x''(t)+0.3 x'(t)-x(t)+(x(t))^2+(x(t))^3=\mu \sin t$. 
Numerical studies show regions of chaos
interspersed with regions without chaos, indicating that our Off-On-Off 
Chaos Theorem applies to this situation. 
\label{Fig:vdp} }
\end{figure}

In Fig.~\ref{Fig:quadratic} as $\mu$ increases towards a value 
$\mu_F \approx 3.57$, one encounters a family of periodic orbits that undergo an infinite sequence of period doublings with the period of these orbits tending to $\infty$.
The appearance of infinitely-many such (period-doubling) cascades is one of
the most prominent features observed in the study of parametrized
maps. Cascades were first reported by Myrberg in
1962~\cite{myrberg:62} ({\it cf.} Fig.~\ref{Fig:quadratic}),
\begin{figure}
  \includegraphics[width=.5\textwidth]{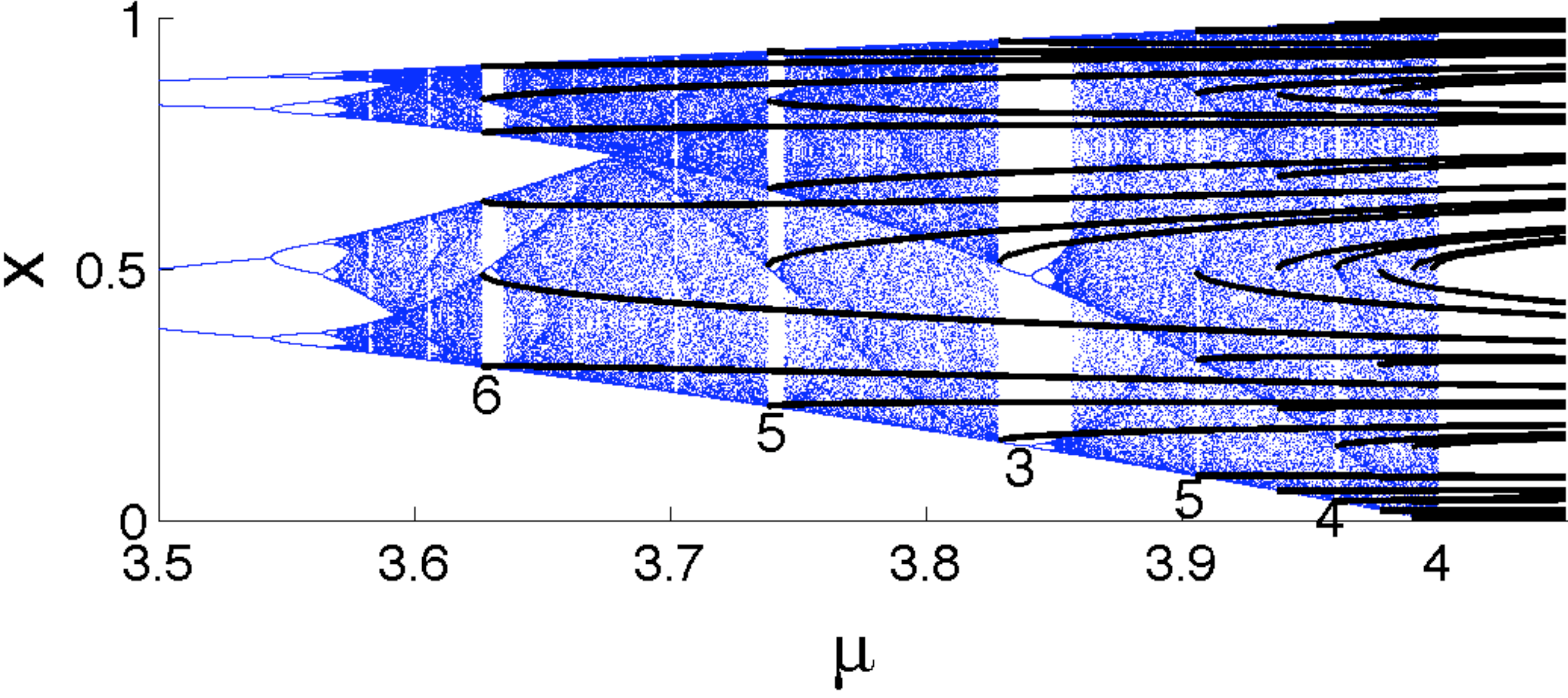}  \label{Fig:quadratic} 
\caption{{\bf Cascades and their connected components.} The attracting set for the
 logistic map $F(\mu,x)=\mu x (1-x)$ is shown in blue. There are
 infinitely-many cascades, each with infinitely-many period-doubling
 bifurcations. Each saddle-node bifurcation creates both a cascade and  a 
path of unstable orbits, the latter shown as black curves for periods up to six. 
 We call the paths of unstable orbits the {\bf stems}
 of the cascades. Below the largest gaps or {\it windows} in the blue regions, we give the period 
 of the associated cascade. For this particular map, the black curves of unstable orbits  extend to $\mu = +\infty$. 
Here, at each sufficiently large
 $\mu$, there is exactly one orbit ($k$ points) in each period-$k$
 stem.
 \label{Fig:quadratic}}
\end{figure}
and Robert May popularized their existence to a huge scientific
audience~\cite{may:74}. They are found in a large variety of
contexts: in Raleigh-Bernard convection~\cite{yahata84}, 
damped bouncing
balls~\cite{tufillaro92},
Van der Pol oscillators~\cite{yu:08} ({\it cf.} Fig.~\ref{Fig:vdp}),
vibratory ball milling~\cite{huang95}, dust charge fluctuation in
plasma systems~\cite{bora07p}, external optical injection in
lasers~\cite{simpson94}, delay oscillators~\cite{larger:etal:05},
vibrating damaged structures~\cite{carpinteri05}, glow
discharge~\cite{sijacic04}, combustion~\cite{frankel94}, very slow
classical Cepheid stars~\cite{buchler00}, neuron and pancreatic
cells~\cite{deng99}, dripping faucets~\cite{gerstecki:etal:05},
bouncing droplets on soap films~\cite{gilet:bush:09},
Belousov-Zhabotinsky reactions~\cite{freire:09}, and in the cromorn, a
medieval musical instrument~\cite{gibiat99}. 

We find that for many
systems depending on a parameter, every periodic orbit is part of a
cascade; for such systems, cascades are as fundamental as periodic orbits
themselves. The scaling properties of individual cascades have been
studied for cascades in nearly quadratic maps~\cite{feigenbaum:79},
but there are only a few results about the existence of
cascades~\cite{yorke:alligood:83}. Furthermore, the mathematical and
scientific literature focuses on the study of single cascades.  In
this Letter, we describe the results of a new general theory of
cascades, which explains why cascades exist and why chaotic dynamical systems
often have infinitely many cascades.

{\bf Two kinds of periodic orbits meet at period-doubling bifurcations.} 
For a map $F(\mu, x)$ that depends on a
parameter $\mu$, 
a point $(\mu, x_0)$ is 
a {\bf period-$p$ point} if 
$F^p(\mu, x_0) = x_0$, where $p > 0$ is chosen as small as possible.
Writing $x_{n+1} = F(\mu, x_n)$, its {\bf (periodic) orbit} is the set of
points $\{(\mu, x_0), (\mu, x_1), (\mu, x_2), \cdots, (\mu,
x_{p-1})\}$.  By the {\bf eigenvalues of} that orbit, we mean the
eigenvalues of the Jacobian matrix $D_xF^p(\mu, x_0)$. If the
map is one-dimensional, then by the derivative of the orbit, we mean
the derivative of $F^p$.  A {\bf bifurcation orbit} is any orbit
having an eigenvalue with absolute value $1$.

We call a periodic orbit a {\bf flip}
orbit if the orbit has an odd number of eigenvalues less than -1, and
-1 is not an eigenvalue.  (In one dimension, this condition is:
derivative $<-1$. In dimension two, flip orbits are those with exactly
one eigenvalue $< -1$.) All other periodic orbits are called {\bf
  regular}. A family switches between flip and regular orbits at a
period-doubling bifurcation orbit since this bifurcation orbit always has an
eigenvalue equal to $-1$.

For $\psi$ in some interval $(a,b)$, let $(\mu(\psi), Y(\psi))$ be a path of
{\it regular} periodic points in $(\mu,y)$ space and write $[Y(\psi)]$
to denote the periodic orbit of the periodic point $Y(\psi)$.  We say
$(\mu(\psi), [Y(\psi)])$ is a {\bf cascade} if the path contains
infinitely many period-doubling bifurcations and has orbits with all
the periods $\{k, 2k, 4k, 8k, \cdots \}$ for some positive integer
$k$.

\begin{figure}
\includegraphics[width=.5\textwidth]{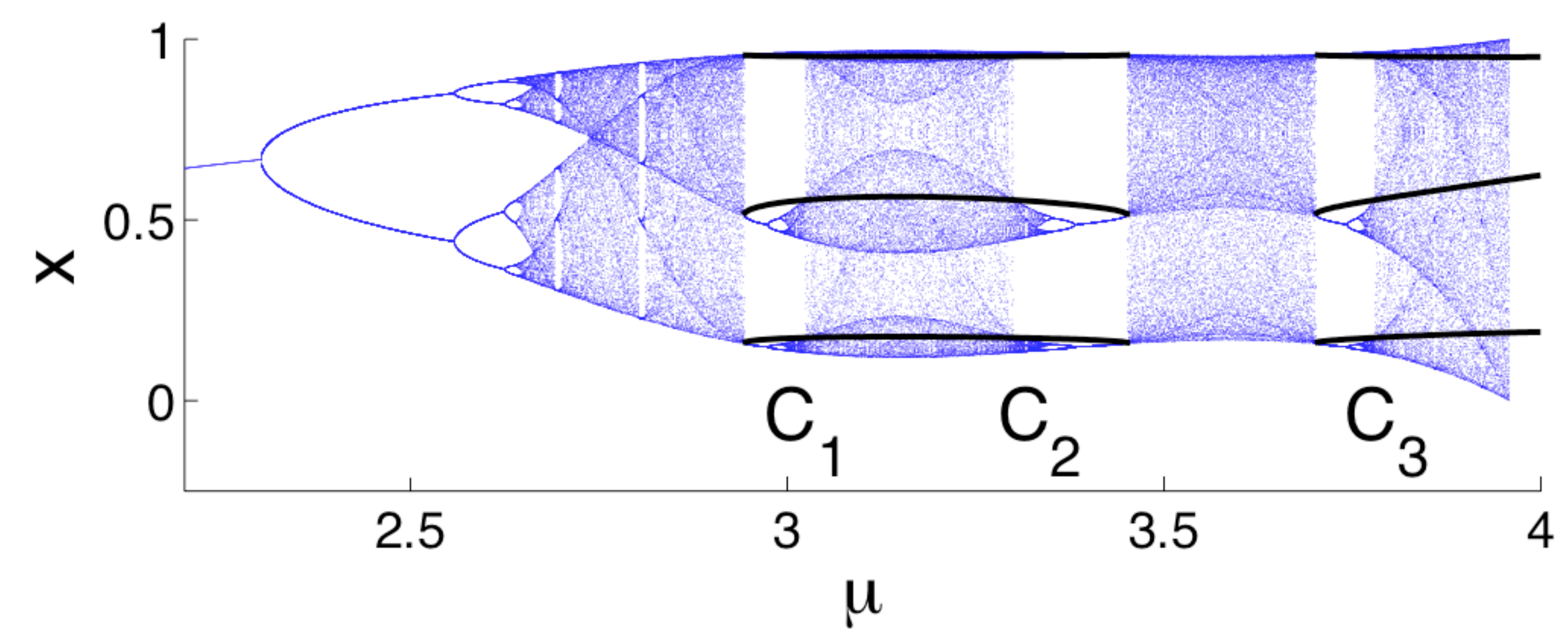}
\caption{{\bf Bounded paired cascades.} 
The attracting set for the function $F(\mu,x)= 
h(\mu)x(1-x)$, where $h(\mu)= \mu (1.18+0.17 \cos(2.4 \mu))$. 
Cascades $C_1$ and $C_2$ constitute a {\bf pair} of cascades, since they are 
connected by a family of orbits (black curves). Cascade $C_3$ is  unbounded and unpaired, 
as indicated by the black curve continuing to the right to $\mu = \infty$. 
\label{Fig:bounded}}
\end{figure}
{\bf Assumptions.} {\it All results in this paper assume that the maps 
$y \mapsto F(\mu,y)$
are infinitely
  differentiable jointly in the parameter $\mu$ and in phase space $y$ and that all of
  the periodic orbit bifurcations are {\bf generic}, 
  meaning that every bifurcation orbit is a standard nondegenerate
  period-doubling, saddle-node, or Hopf bifurcation.} For example, generic maps 
 have neither symmetries nor symmetry breaking bifurcations. Nor do they
have pitchfork or period-tripling bifurcations.  We
also assume that for any parameter interval $[\mu_1, \mu_2]$, all periodic orbits are contained in a bounded set in
phase space.
 
{\bf The cascades route to chaos in dimensions one and two.}  Usually the term
``route to chaos'' refers to formation of {\it chaotic attractors}, in
which case there are several possible routes to chaos.  In this
Letter, we consider routes from a parameter having no chaos to one with
{\it chaotic sets}, where the sets are not necessarily attractors.  In
fact, we require only one aspect of chaos: we say that there is {\bf chaos} at a particular $\mu$ if there are infinitely
many regular periodic orbits. 
For example, there is chaos whenever there is a transverse homoclinic point. That is equivalent to
having a horseshoe for some iterate of the map, and it implies there are
infinitely many regular saddles in two dimensions, and infinitely many
regular unstable orbits in dimension one.  This definition of chaos is
sufficiently general as to include having one or multiple coexisting
chaotic attractors, as well as the case of transient chaos. We say a
map has {\bf no chaos} at a particular $\mu$ if there are at most
finitely many regular periodic orbits.

{\bf Cascades Route to Chaos Theorem.}  {\em Assume the parametrized
  map $F(\mu,\cdot)$ has a one- or two-dimensional phase space. If at
  parameter value $\mu_1$ there is no chaos, and at $\mu_2$, there is
  chaos and there are at most finitely many attracting orbits 
  (and finitely many repelling
  orbits in dimension 2), then $F$ has infinitely many cascades
  between $\mu_1$ and $\mu_2$. } 

Thus for one- and two- dimensional smooth families,
the only route to chaos is through infinitely many cascades. It is for
instance impossible to get chaos at $\mu_2$ from a single cascade.

If we were to omit the assumption that there are infinitely many {\it
  regular} orbits, the conclusion would be false. For example, in the
quadratic map let $\mu_F$ denote the Feigenbaum parameter value, i.e., the first
parameter value where the period doublings accumulate. At $\mu_F$
there are infinitely many periodic orbits, and all but one 
are flip orbits. Furthermore $\mu_F$ is preceded by a single
cascade.

Each cascade will have some minimum period $p$ and will have periodic
orbits of periods $p, 2p, 4p, 8p, \cdots$. If the map is one-dimensional 
or is two-dimensional and dissipative in the sense that
there are no repelling periodic orbits, then 
each of these periods
is the period of some attracting orbits in the path, though not all
orbits will necessarily be attracting.

\begin{figure}
\includegraphics[width=.5\textwidth]{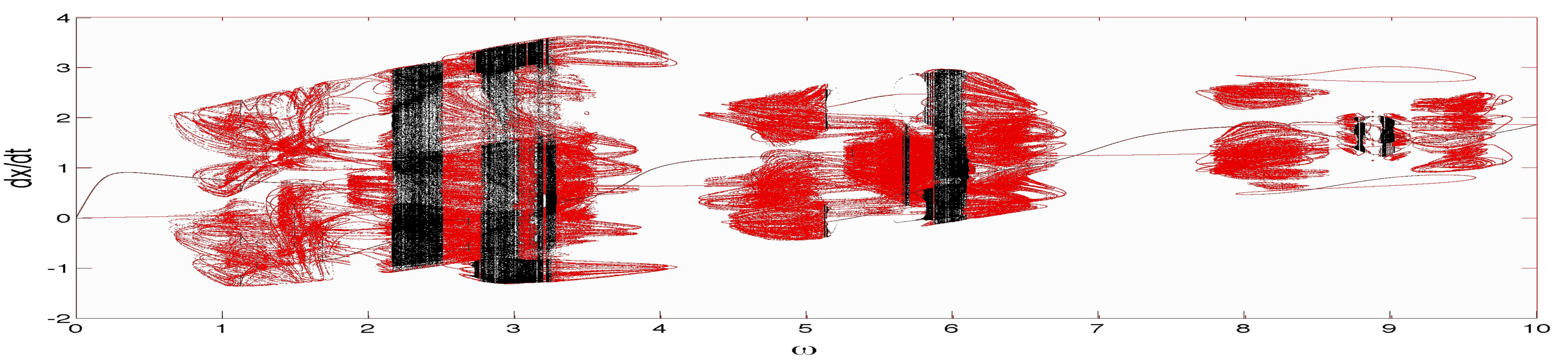}
\caption{{\bf The forced-damped pendulum.} 
For this figure, periodic points with periods $< 10$ were plotted 
in red for the time-$2\pi$ map of the forced-damped pendulum equation: 
$x''(t)+0.2 x'(t)+\sin(x(t))=\omega \cos(t)$, indicating the general areas with chaotic dynamics for this map.
Then the attracting sets were plotted in black, hiding some periodic points. 
Much more detailed calculations confirm that interspersed with the chaos, 
there are  some parameter ranges without chaos. 
\label{Fig:pend} }
\end{figure}

Based on numerical
studies, a number of maps appear to satisfy the conditions of the
Cascades Route to Chaos Theorem. Note that these numerical verifications
involve significantly more work than just plotting the attracting sets
for each parameter, since we are concerned about both the stable and
the unstable behavior to determine whether there is chaos. Examples
include the time-$2\pi$ maps for the double-well Duffing (Fig.~\ref{Fig:vdp}), 
the triple-well Duffing, and forced-damped pendulum (Fig.~\ref{Fig:pend}), 
as well as the Ikeda map used to describe the
field of a laser cavity, and the pulsed damped rotor map.

We now discuss the creation of a path $(\mu(\psi),[Y(\psi)])$ of {\it
  regular orbits}. Consider the procedure of starting at a regular
periodic orbit and following the path of periodic orbits containing it
in $(\mu, y)$-space, following only regular orbits. Follow the path --
such as numerically -- through saddle-node bifurcations by reversing
the direction in the parameter space. When period-$p$ orbits reach a
period-doubling bifurcation $(\mu_0,[Y])$, follow the period-$2p$ orbits. They
are regular orbits, whereas the extension past $(\mu_0,[Y])$ of period-$p$
orbits are flip orbits.  Likewise, at a period-halving bifurcation,
follow along the branch of regular orbits.  Following a path away from
a cascade can lead to $|\mu| \to \infty$ or it can lead to a second
cascade. We call an entire connected path of regular orbits a {\bf
  component}.  Two orbits are in the same component if and only if it
is possible to follow a path of regular orbits from one orbit to the
other continuously, without jumps, though the period is allowed to
double many times.

{\bf There are two types of cascades.} Only one type of cascades
occurs for quadratic maps, but there is a second type of cascade as
well. Namely, the component of a cascade either lies in a bounded
region of $(\mu, y)$-space, or it does not. Accordingly we call the
cascade {\bf bounded} or {\bf unbounded}, respectively, though the
adjective really describes the cascade's component.  Each cascade in
the quadratic map ({\it cf.} Fig.~\ref{Fig:quadratic}) is unbounded because
a branch of periodic orbits extends from each cascade to $\mu =
\infty$. The constant period branch that extends to infinity is called
the {\bf stem} of the (unbounded) cascade.  Thus the theory of
quadratic maps is the theory of unbounded cascades.  We call the
period of the orbits in the stem, the {\bf stem period of the
  unbounded cascade}.  (We only treat the case of an unbounded cascade
which has a constant period stem.)  

We refer to two cascades as a {\bf
  pair} if they are in the same component; that is, if the two
cascades have a path of periodic orbits running from one the other
({\it cf.}  Fig.~\ref{Fig:bounded}).  We have shown the following rigorous
result.

{\bf Bounded Paired Cascades Theorem.} {\em Bounded cascades
always come in pairs.}

The theorem below shows that paired cascades are common. Once again we consider $x$ in a one- or
two-dimensional phase space. We say that map $F$ has {\bf off-on-off
chaos} for $\mu_1<\mu_2<\mu_3$ if there is no chaos at
$\mu_1$ and $\mu_3$, whereas at $\mu_2$, $F$ is as assumed in the Cascades
Route to Chaos Theorem. 
 
{\bf Off-On-Off Chaos Theorem.} {\em If $F$ is two-dimensional and has
off-on-off chaos for $\mu_1<\mu_2<\mu_3$, then $F$ has infinitely many
bounded paired cascades and at most finitely many unbounded cascades.} 

Our numerical studies indicate that the
time-$2\pi$ maps of the forced double-well Duffing (Fig.~\ref{Fig:vdp}) and forced damped
pendulum (Fig.~\ref{Fig:pend}) have off-on-off chaos, and that it occurs on multiple
non-overlapping parameter regions. Here $2\pi$ is the forcing period. In fact we cannot prove there are
only finitely many periodic attractors, though we find very few. These systems
have no periodic repellers.

{\bf Conservation of unbounded cascades.}  If in addition to the
hypotheses of the Cascades Route to Chaos Theorem, we assume further that ({\it i}) at
$\mu_1$ there are no regular periodic orbits and ({\it ii}) at $\mu_2$ there are
no attractors (nor repellers in dimension two), then the component of
{\it each} regular orbit at $\mu_2$ contains a cascade that is between
$\mu_1$ and $\mu_2$, and distinct regular orbits at $\mu_2$ are in different
components. Starting at each regular orbit at $\mu_2$, there is a 
unique path of regular orbits -- initially in the
direction of $\mu_1$ -- ending in a cascade. 
Thus the number of cascades hitting the boundary of the parameter interval
depends only on the number of regular orbits at $\mu_2$, independent
of the behavior of the map between $\mu_1$ and $\mu_2$.  

This leads to a heuristic {\bf conservation principle for unbounded cascades}: 
if $F(\mu,x)$ satisfies property ({\it i}) for $\mu$ sufficiently negative and
property ({\it ii}) for $\mu$ sufficiently positive, then any perturbation
$F(\mu,x)+g(\mu,x)$ such that $F$ dominates $g$ for $|\mu| \to \infty$  has
the same set of unbounded cascades.

Three rigorous examples of this idea are encapsulated in the following 
one-dimensional maps 
\begin{eqnarray}\label{eqn:quad+g}
F(\mu,x) = \mu - x^2 +g(\mu,x) &(\mbox{quadratic}),\\
F(\mu,x)= \mu x-x^3  + g(\mu,x) &(\mbox{cubic}),\nonumber \\
F(\mu,x)=x^4- 2\mu x^2 + {\mu^2}/{2}+ g(\mu,x) &(\mbox{quartic}), \nonumber
\end{eqnarray}
where for some real positive $\beta$, 
\begin{eqnarray}\label{eqn:g1} 
|g(\mu,0)| < \beta &\mbox{ for all } \mu, \mbox{ and } \\
|g_x(\mu,x)| < \beta &\mbox{ for all } \mu,x. \nonumber 
\end{eqnarray}
For such $g$, each of the three maps each has no regular periodic orbits 
for $\mu$ sufficiently negative, and for $\mu$ sufficiently large has a
one-dimensional horseshoe map. 
The quartic map was chosen so that when $g$ is identically 0 and $\mu > 0$,
the graph has two minima, at $(\pm \sqrt{\mu}, -\mu^2/2)$, and the
local maximum is at $(0, +\mu^2/2)$.   The
conditions on $g$ guarantee that it does not significantly affect the
periodic orbits when $|\mu|$ is sufficiently large; in particular it
does not affect their eigenvalues, so it does not affect the number of
period-$p$ regular periodic orbits for large $|\mu|$.  
Hence one can
check that all three maps have no regular periodic orbits
for $\mu$ very negative and have no attracting periodic orbits for $\mu$ very positive, and for sufficiently large $|\mu|$, all periodic orbits are contained in the set $[-2 \sqrt{\mu},2 \sqrt{\mu}]$. 
Thus the following holds.

{\bf Conservation of cascades theorem.}  
{\em For any $F$ of the
functions in Eqn.~\ref{eqn:quad+g}, with $g$ chosen as in
Eqn.~\ref{eqn:g1}, the number of stem-period-$k$ unbounded cascades
is independent of the choice of g. }

\begin{table}[t]\label{table1}
  \begin{tabular}{ | l | l | l |l|}
  \hline
  $k$ & Quad$(k)$ & Cubic$(k)$ & Quart$(k)$ \\ \hline
  1& 1 & 2 & 2\\ \hline
  2& 0 & 1 & 2\\ \hline
  3& 1 & 4& 10\\ \hline
  4& 1 & 8& 28\\ \hline
  5&	3 & 24 & 102\\ \hline
  6&	4 & 56 & 330\\ \hline
  7&	9 & 156& 1152\\ \hline
  8&	14 & 400 & 4064\\ \hline
  9&	28 & 1092& 14560\\ \hline
  $k \gg 10$& $\displaystyle \sim 2^k/2k$ & $\displaystyle \sim 3^k/2k$ & $\displaystyle \sim 4^k/2k$ \\ \hline
  \end{tabular}
  \caption{
   The number of unbounded cascades of stem period $k$ for any large-scale 
   perturbation of the quadratic, cubic, and quartic maps, respectively 
   labeled $Quad(k)$, $Cubic(k)$, and $Quart(k)$.}
\end{table}

The number of unbounded cascades in each of these three cases is
summarized in our Table.  The relative sizes reflect the complexity of
the three horseshoe maps; the quadratic map has the complexity of a
standard two-branch horseshoe, whereas the cubic has a horseshoe with
three branches, and the quartic has four branches.

The conservation principle works for two- and higher-dimensional maps
as well. We have shown in~\cite{sander:yorke:p09b} that large-scale
perturbations of the two-dimensional H{\'e}non map conserve unbounded
cascades.  In~\cite{sander:yorke:p09}, we have shown that there is
conservation of unbounded cascades for a coupled system of $N$
quadratic maps.

\section{Acknowledgements}
Thank you to Safa Motesharrei for his corrections and comments. 
J.A.Y. was partially supported by NSF Grant DMS-0616585 and NIH Grant
R01-HG0294501. E.S. was partially supported by NSF Grants DMS-0639300 and DMS-0907818,
and NIH Grant R01-MH79502.

\begin{flushleft}
%
%
\addcontentsline{toc}{subsection}{References}
\footnotesize
%
%
\bibliographystyle{aps}

\end{flushleft}

\end{document}